\documentclass[a4paper,11pt]{article}
\pdfoutput=1 

\usepackage{jcappub} 

\usepackage[T1]{fontenc} 
\usepackage{xcolor}
\usepackage{graphicx}
\usepackage{isotope}
\usepackage{subcaption}
\usepackage{array}
\usepackage{amsmath}
\usepackage{hyperref}

\newcommand{\lya}{Ly$\alpha$}
\newcommand{\lyaf}{Ly$\alpha$ forest}
\newcommand{\lyb}{Ly$\beta$}
\newcommand{\lyalya}{Ly$\alpha \times$Ly$\alpha$}
\newcommand{\lyaqso}{Ly$\alpha \times$QSO}
\newcommand{\chisq}{$\chi^2$}

\newcommand{\apar}{$\alpha_{||}$}
\newcommand{\aperp}{$\alpha_\bot$}

\title{\boldmath Bayesian methods for fitting Baryon Acoustic Oscillations in the Lyman-$\alpha$ forest}

\author[a]{Andrei Cuceu}
\author[a]{Andreu Font-Ribera}
\author[a]{Benjamin Joachimi}

\affiliation[a]{Department of Physics and Astronomy, University College London, London, UK}

\emailAdd{andrei.cuceu.14@ucl.ac.uk}
\emailAdd{a.font@ucl.ac.uk}
\emailAdd{b.joachimi@ucl.ac.uk}

\abstract{We study and compare fitting methods for the Lyman-$\alpha$ (\lya) forest 3D correlation function. We use the nested sampler \texttt{PolyChord} and the community code \texttt{picca} to perform a Bayesian analysis which we compare with previous frequentist analyses. By studying synthetic correlation functions, we find that the frequentist profile likelihood produces results in good agreement with a full Bayesian analysis. On the other hand, Maximum Likelihood Estimation with the Gaussian approximation for the uncertainties is inadequate for current data sets. We compute for the first time the full posterior distribution from the \lyaf\ correlation functions measured by the extended Baryon Oscillation Spectroscopic Survey (eBOSS). We highlight the benefits of sampling the full posterior distribution by expanding the baseline analysis to better understand the contamination by Damped \lya\ systems (DLAs). We make our improvements and results publicly available as part of the \texttt{picca} package.}

\begin{document}
\maketitle
\flushbottom


\section{Introduction}
\label{sec:intro}

Over the last few decades, cosmology has entered a data driven era, with large surveys providing rich data sets. These data sets encode a vast amount of information, which most of the time is non-trivial to extract. Performing such large surveys is very expensive and time consuming, which puts even more emphasis on efficient and accurate extraction of meaningful information. Over the last two decades, the $\Lambda$ Cold Dark Matter ($\Lambda$CDM) model has become widely accepted as the standard cosmological model, however, it has 6 free parameters with extensions adding even more. On top of this, most analyses need extra nuisance parameters to create good models of their data, leading to very high-dimensional parameter spaces (e.g. \cite{PlanckCollaboration:2018,Abbott:2018}). It has become essential to have reliable analysis tools, and this has lead to an increased focus on statistical methods and interpretation.

The efficient and reliable extraction of Baryon Acoustic Oscillation (BAO) information from large scale structure (LSS) data has been a very active topic of research over the last 15 years. Since the first detections of BAO, using the distribution of galaxies \cite{Eisenstein:2005,Cole:2005}, there has been much attention given to optimising statistical methods used on data of discrete tracers, e.g. \cite{Vargas:2013,Anderson:2014,Chan:2018,Hinton:2020}. However, the newer method of measuring BAO using the Lyman-$\alpha$ (\lya) forest has received comparatively less attention. 

The first detection of the BAO scale in the \lyaf\ auto-correlation (\lyalya) function was done by the Baryon Oscillation Spectroscopic Survey (BOSS) using Data Release 9 (DR9) of the Sloan Digital Sky Survey (SDSS) \cite{Busca:2013,Slosar:2013,Kirkby:2013}. It was also detected in the cross-correlation of \lya\ absorbers with quasar positions (\lyaqso) using BOSS DR11 \cite{FontRibera:2014}. A physical model for the contaminations was first used by \cite{Bautista:2017,duMasdesBourboux:2017}, however it has a large number of parameters that model the contamination by Damped \lya\ systems (DLAs) and different metal absorption lines. Recent analyses have introduced yet more effects \cite{DeSainteAgathe:2019,Blomqvist:2019}, and these have led to a large parameter space and potentially complex behaviour. However, as the main aim has always been the measurement of the BAO peak position, these astrophysical parameters have received little attention.

The BAO scale can be measured using just two parameters: \apar\ and \aperp. These measure the size of the BAO scale relative to a fiducial cosmology along and across the line of sight, respectively. Because of the focus on measuring BAO, the high-dimensional parameter spaces have so far been investigated only using a frequentist methodology. In particular, the profile likelihood (e.g. \cite{Planck:2014,Chan:2018,Algeri:2019}) has been used to extract the relevant information. This method approximates the probability at some value of \apar\ and \aperp\ by the maximum likelihood over the nuisance parameters at that point. In this work we use a Bayesian approach to fit the \lyaf\ 3D correlation function and we provide a tool for studying this large parameter space and for extracting all the relevant information. While Bayesian methods have been used in \lya\ forest analyses before (e.g. \cite{Pichon:2001,Kitaura:2012,Horowitz:2019,Porqueres:2019}), they were never used in BAO analyses of the \lya\ correlation function.

The principal difference between the Bayesian \cite{Bayes:1763,Laplace:1820} and frequentist \cite{Neyman:1937} methodologies is their interpretation of the concept of probability. In the Bayesian framework, probability is a degree of belief in an event, while in the frequentist framework, the probability of an event is the limit of its relative frequency in many trials. In the limit of infinite data, the two approaches produce the same results. Our focus, however, is not on philosophical interpretations, but on the practical consequences of the two frameworks when working with real data. One of the reasons for the widespread use of Bayesian methods is the availability of samplers such as Monte Carlo Markov Chain (MCMC) \cite{Metropolis:1953,Hastings:1970} which facilitate the efficient exploration of the complex high-dimensional posterior distributions that often appear in cosmology. It is the efficiency and accuracy of such tools that we want to compare with equivalent frequentist approaches within the context of fitting the \lyaf\ correlation function.

The purpose of this work is to investigate the methods used so far and compare them with a Bayesian framework. In particular, we use for the first time a sampler to obtain the full posterior distribution of all parameters. We begin in Section \ref{sec:theory}, where we discuss the frequentist methods used so far, and compare them with the Bayesian methodology. In Section \ref{sec:mocks}, we use synthetic correlation functions to showcase the similarities and differences of the two approaches when fitting the BAO parameters. Finally, in Section \ref{sec:data}, we use a Bayesian framework to analyse the latest extended BOSS (eBOSS) DR14 correlation functions \cite{DeSainteAgathe:2019,Blomqvist:2019}, and highlight some potential uses and advantages of the full posterior distribution.

\section{Bayesian vs. frequentist methods}
\label{sec:theory}

We first discuss some of the theoretical differences between the Bayesian and frequentist approaches to statistical data analysis. We include this discussion for completeness and also because, as we will see, the two frameworks answer fundamentally different questions when it comes to the quantification of uncertainties. As such, we believe that these theoretical considerations are important to the interpretation of our results. We also discuss some practical differences and their implications for cosmology and we conclude with a simple toy example.

\subsection{The best fit model}

The difference in the interpretation of probability between the Bayesian and frequentist approaches leads to a difference in their principal object of study. The frequentist approach treats parameters as fixed quantities and the data as the random variable. The object of study is the probability of obtaining the data, $D$, given a model, $\mathcal{M}$, and some parameters, $\vec{\theta} = (\theta_1,...,\theta_n)$; this is also known as the likelihood: $P(D|\vec{\theta},\mathcal{M})$. For normally distributed data the likelihood takes the form:
\begin{equation}
    P(D|\vec{\theta},\mathcal{M}) = \frac{\exp \Big[- \frac{1}{2} \big(D - \mathcal{M}(\vec{\theta})\big)^T \Sigma^{-1} \big(D - \mathcal{M}(\vec{\theta}) \big)\Big]}{\sqrt{(2\pi)^n |\Sigma|}},
    \label{eq:lik}
\end{equation}
where $\Sigma$ is the covariance matrix of the data. We will also refer to the logarithm of the likelihood which we denote by $\mathcal{L} \equiv \log P(D|\vec{\theta},\mathcal{M})$.

The object of interest for a Bayesian is the posterior distribution of the parameters $\vec{\theta}$, given the data, and a model: $P(\vec{\theta}|D,\mathcal{M})$. This fully encapsulates our knowledge of the probability of possible values of the parameters of interest, by treating these parameters as random variables. The posterior distribution can be computed through Bayes' Theorem \cite{Bayes:1763,Laplace:1820}:
\begin{equation}
    P(\vec{\theta}|D,\mathcal{M}) = \frac{P(D|\vec{\theta},\mathcal{M}) P(\vec{\theta}|\mathcal{M})}{P(D|\mathcal{M})},
    \label{eq:bayes}
\end{equation}
where $P(\vec{\theta}|\mathcal{M})$ is the prior probability, and $P(D|\mathcal{M})$ is a constant (for a model $\mathcal{M}$) known as the Bayesian evidence. The evidence is the normalization of the posterior and requires an n-dimensional integral to be computed. It can be used to perform Bayesian model selection (see e.g. \cite{Liddle:2006,Trotta:2007,Trotta:2008} for applications in cosmology), however, when the only goal is inference this quantity is not necessary.

If we work with wide flat priors, we can deduce from Equations \ref{eq:lik} and \ref{eq:bayes} that the two frameworks will produce the same best fit $\vec{\theta}_{best}$, given by the frequentist maximum likelihood $P_{max}(D|\vec{\theta},\mathcal{M})$, and by the Bayesian maximum posterior probability $P_{max}(\vec{\theta}|D,\mathcal{M})$. 



\subsection{Quantifying uncertainties}
\label{subsec:uncert}

The two approaches diverge again when it comes to finding the uncertainty on $\vec{\theta}_{best}$. In this case it is not only a difference in methodology, but also a fundamental difference in the object of interest. Frequentists quantify uncertainty through confidence intervals\footnote{For n-dimensional distributions these are referred to as confidence regions, however we chose to only talk about intervals to clearly distinguish them from Bayesian credible regions} (CI), which are defined by the proportion (frequency) of intervals, measured from the ensemble of possible data sets, that contain the true values of the parameters ($\vec{\theta}_{true}$). On the other hand, Bayesian uncertainty is quantified through credible regions (CR), defined as the smallest region of the posterior that encompasses a certain probability (most often $68\%$ and $95\%$ CRs are quoted). 

The two questions asked by Bayesians and frequentists are very different. A Bayesian CR is telling us that, given our data and prior, we are e.g. $95\%$ confident that the true values $\vec{\theta}_{true}$ are within that region. Meanwhile, the frequentist CI is telling us that if we repeat our experiment many times, the confidence intervals we obtain will contain the true values $\vec{\theta}_{true}$ in e.g. $95\%$ of the cases. Note that the frequentist CI that we obtain from our data does not state anything about the probability that it contains $\vec{\theta}_{true}$ (this is a common misconception). In fact, there are extreme cases in the literature where a frequentist CI has $0\%$ probability of containing the truth\footnote{We must stress however, that even in such extreme cases the frequentist CIs are not wrong. They are just answering a different question.} \cite{Jaynes:1976,Welch:1939}. This behaviour is due to the fact that in the frequentist methodology, one never conditions a result on the actually observed data $D$, but instead on its distribution of possible realisations, which we do not always fully understand given limited data. This issue is fundamental in cosmology, because we only have one Universe to observe. Bayesian CRs can also cause problems because of prior choices. For example, flat priors are only flat for the specific parametrisation they are defined on. 

We now turn to the computation of these uncertainty intervals and regions. In a frequentist framework, we usually start by computing the best fit parameters $\vec{\theta}_{best}$ by maximising the likelihood. This is called Maximum Likelihood Estimation (MLE). After that, we can compute a covariance matrix for the parameters by taking the second derivative of the likelihood in parameter space around this peak. However, this covariance matrix is only accurate in general if the errors on the data are normally distributed and the model is linear in all parameters (which would correspond to a Gaussian posterior in a Bayesian framework). If this is not the case, it can still be applied around the peak, but a better approach is to compute the likelihood on a grid in parameter space (we will call this a scan). 

The confidence intervals can be computed using regions of equal likelihood around $\vec{\theta}_{best}$. To find these regions we need $\Delta\mathcal{L}_p \equiv \mathcal{L}_p - \mathcal{L}_{min}$ values such that the region defined by $\mathcal{L}_p$ corresponds to a certain CI of probability $p$. In the Gaussian case these values can be computed analytically. In the general case, a large number of Monte Carlo simulations of the data are needed to compute $\Delta\mathcal{L}_p$ values. This procedure is outlined in Appendix \ref{sec:app_chisq}. Using these $\Delta\mathcal{L}_p$ values, frequentists can draw constant likelihood contours using a scan of the parameters and obtain the correct confidence intervals.

We must stress that from a Bayesian perspective the best fit and the credible regions are just special values computed from the posterior. It is this full posterior distribution that is the real object of interest because it contains all the information about the probable values of the parameters. The computation of the posterior distribution can however be a very demanding task. When dealing with low-dimensional spaces, Bayesians can compute the posterior on a grid, similar to the frequentist method. The scan can be used to compute the Bayesian credible regions by finding the smallest region of the scan that contains a certain probability $p$. This is usually done by ordering the scan in decreasing order of probability and computing their running sum until the result is a fraction $p$ of the total probability of the grid:
\begin{equation}
    \sum_{j=1}^M P(\vec{\theta}_j|D,\mathcal{M})  = p \sum_{j=1}^N P(\vec{\theta}_j|D,\mathcal{M}),
    \label{eq:bayes_scan}
\end{equation}
where $N$ is the total number of points on the grid, and the $M$ points obtained through this method cover the CR of probability $p$. Note that the integral of the posterior over a region is normally required to do this, however, this can be approximated by a sum if the grid is equally spaced because the probability density at each point is proportional to the probability mass for that region. As we will show in the next Section, these two interpretations of the same grid results often produce identical results. However, the scan quickly becomes infeasible with increasing number of parameters, and Bayesians move on to using more effective methods, such as MCMC.

Bayesians can deal with high-dimensional spaces by efficient and accurate sampling, using tools such as MCMC \cite{Metropolis:1953,Hastings:1970} (see e.g. \cite{Press:2007,Gelman:2013} for detailed introductions). Furthermore, the efficient computation of the Bayesian evidence has also become possible with the introduction of Nested Sampling \cite{Skilling:2004}. The underlying principle of such methods is the creation of samples from continuous random variables with probability density proportional to a known function, in our case the unnormalized posterior distribution. Once a sufficient number of samples have been generated, they can be used to compute summary statistics. For example, computing the credible regions amounts to finding the smallest region containing a certain fraction of samples. This fraction corresponds to the probability that, given the data and the prior, the true values of the parameters are within that region.


\subsection{Nuisance parameters}

The handling of nuisance parameters is also an intense topic of debate. In a Bayesian framework, the answer is marginalization. If the parameter vector contains two sets: interesting parameters $\vec{\theta}_\textrm{i}$ and nuisance parameters $\vec{\theta}_\textrm{n}$, then the posterior distribution of the interesting parameters is given by:
\begin{equation}
    P(\vec{\theta}_\textrm{i}|D,\mathcal{M}) = \int_{\vec{\theta}_\textrm{n}} P(\vec{\theta}_\textrm{i},\vec{\theta}_\textrm{n}|D,\mathcal{M}) \; d\vec{\theta}_\textrm{n}.
\end{equation}
We partition the full posterior by integrating over the nuisance parameter space. This ensures that all of the probability mass contained in the nuisance parameters is accounted for. On the other hand, nuisance parameters are a major problem of non-Bayesian statistical theories \cite{Ghosh:1988}. There is no consensus frequentist way of addressing this problem (see \cite{Basu:1977} for a discussion of some of the methods). 

One of the most common frequentist methods is the profile likelihood (e.g. \cite{Planck:2014,Chan:2018,Algeri:2019}), which involves computing the likelihood of $\vec{\theta}_\textrm{i}$ on a grid, while conditioning on special values of the nuisance parameters, e.g. the best fit values $\vec{\theta}_{\textrm{n},best}$. In practice, this is done by maximising the likelihood over all nuisance parameters at every point on the grid of interesting parameters:
\begin{equation}
    P(D|\vec{\theta}_\textrm{i},\mathcal{M}) \propto \max_{\vec{\theta}_\textrm{n}} P(D|\vec{\theta}_\textrm{i},\vec{\theta}_\textrm{n},\mathcal{M}).
\end{equation}
For a Gaussian distribution this is equivalent to marginalization because low dimensional cuts from a high dimensional Gaussian are also Gaussian and their volume scales exactly the same as their peak. However, this method can go wrong when a lower dimensional cut changes shape depending on the position of the cut (we show an example of this in Section \ref{subsec:toy}), because it effectively ignores any possibility that the nuisance parameters can have other values. 

The profile likelihood is also useful for computing confidence levels when there are many interesting parameters. The scanning method presented in Section \ref{subsec:uncert} becomes infeasible if we have too many parameters, and as such, the recursive application of the profile likelihood can be a useful approximation. We can scan a small subset of the parameters (usually one or two at a time) and treat the other parameters as nuisance by applying the profile likelihood at every point on the grid. If we apply this method recursively, we can compute an approximation of the full CIs. The profile likelihood is the method used so far in analyses of the \lyaf\ correlation function by BOSS and eBOSS \cite{Slosar:2013,Delubac:2015,FontRibera:2014,Bautista:2017,duMasdesBourboux:2017,Blomqvist:2019,DeSainteAgathe:2019}.

\subsection{Toy Example}
\label{subsec:toy}

\begin{figure}
    \centering
    \includegraphics[width=1.0\textwidth,keepaspectratio]{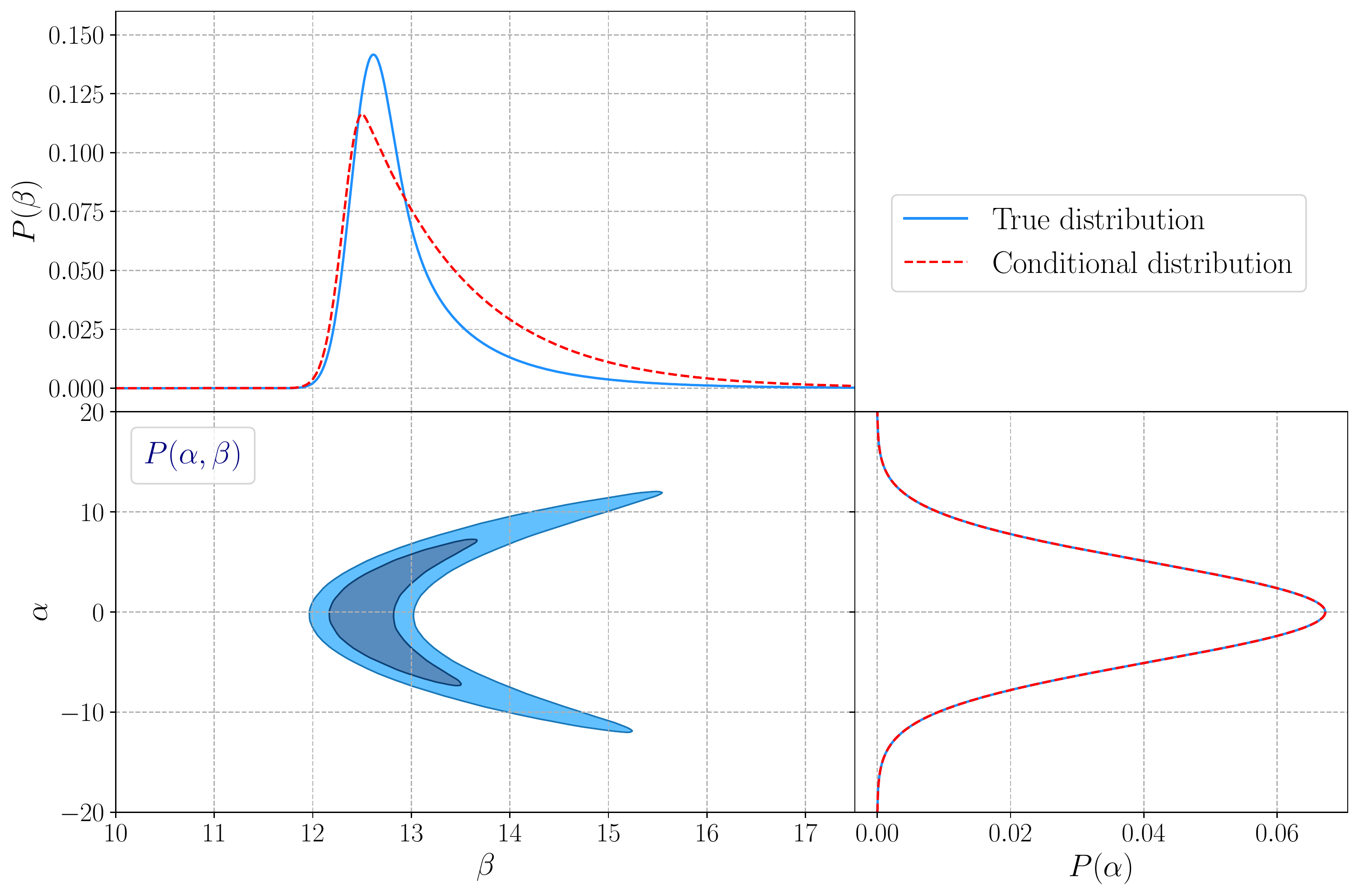}
    \caption{Toy model of a bivariate distribution $P(\alpha,\beta)$ used to illustrate the profile likelihood, a common frequentist approximation for dealing with high-dimensional problems. A scan over $\alpha$ conditional on the best fit value of $\beta$ at each point gives identical results to the marginal distribution (right panel). On the other hand, an equivalent scan over $\beta$ gives the wrong result (top panel), because the conditional distribution $P(\alpha|\beta)$ changes shape depending on the value of $\beta$.}
    \label{fig:toy}
\end{figure}

We illustrate the profile likelihood using a toy example in Figure \ref{fig:toy}. We chose a bivariate distribution $P(\alpha,\beta)$ used in \cite{Wang:2017ex}, which, depending on the parameter of interest, shows where the profile likelihood works perfectly as well as where it fails. If the interesting parameter is ${\theta}_\textrm{i} = \alpha$, a scan over $\alpha$ conditional on the best fit value of $\beta$ at each point gives a result identical to the marginal distribution (bottom right panel of Figure \ref{fig:toy}). On the other hand, if ${\theta}_\textrm{i} = \beta$, the equivalent procedure gives the wrong result because on the right hand side of the $\beta$ grid, the conditional distribution $P(\alpha|\beta)$ is multimodal. This means the shape of $P(\alpha|\beta)$ changes depending on the value of $\beta$, resulting in a failure of the profile likelihood approximation.

This behaviour is not due to the difference in interpretation between a Bayesian and a frequentist. In fact, for this bivariate case, a simultaneous scan of both parameters would produce frequentist results identical to the Bayesian ones, but this is infeasible in high dimensional spaces. If such pathological cases are correctly identified, frequentists can rely on reparameterizations to Gaussianize the problem. However, such cases could appear over arbitrarily many dimensions and be easily missed because usually we can only investigate 1 and 2 dimensional projections.

\section{Testing BAO measurements on mock data sets}
\label{sec:mocks}

We begin our investigation by applying the different methods introduced above to the problem of efficient and accurate extraction of BAO information from the \lyaf\ correlation function. To this end, we use a baseline model of the correlation function to create synthetic data sets. The baseline model is based on the \lya\ correlation functions measured by the extended Baryon Oscillation Spectroscopic Survey (eBOSS) using SDSS DR14 data \cite{DeSainteAgathe:2019,Blomqvist:2019}, and the publicly available modelling package \texttt{picca}\footnote{Available at \url{https://github.com/igmhub/picca}}. Their main analysis follows the frequentist methods introduced in Section \ref{sec:theory}. We perform a Bayesian analysis of the synthetic data sets and compare the different methodologies, with a focus on accurate BAO measurement.

\subsection{Synthetic Correlation Functions}

In order to compare different fitting methods, we produce 100 Monte Carlo simulations of the \lyaf\ flux 3D correlation function. This allows us to investigate the differences over the entire population of possible correlation functions given a data set such as SDSS DR14. For simplicity, in this section we focus only on the \lyalya\ auto-correlation function using \lya\ absorbers only in the \lya\ region, and leave the analysis using the full DR14 data for the next section. We use the measured \lya\ correlation function from SDSS DR14 (shown in Figure 8 of \cite{DeSainteAgathe:2019}). We fit this using the full model from \cite{DeSainteAgathe:2019}, including metal contaminations.

The mock data sets are drawn randomly from a multivariate normal distribution with mean $\xi(\vec{\theta}_{best})$ and covariance $C$, where $\xi(\vec{\theta})$ is the best fit model of the correlation function measured in the DR14 analysis \cite{DeSainteAgathe:2019} and $C$ is the covariance matrix of $\xi$. A new simulated correlation function is then given by:
\begin{equation}
    \tilde{\xi} = \xi(\vec{\theta}_{best}) + A \vec{y},
\end{equation}
where the matrix A comes from the Cholesky decomposition $C = AA^T$, and $\vec{y}$ is a vector of N (the size of the $C$) independent standard normal variates.

The model used to fit the \lyalya\ correlation function in the DR14 analyses has 11 free parameters, of which nine are considered nuisance parameters and two (\apar, \aperp) are the parameters of interest \cite{DeSainteAgathe:2019}. We follow this distinction and leave the analysis and description of the nuisance parameters for the next section. We fit our mocks using this model and the four methods introduced above. We use large flat priors for most parameters, but we follow \cite{DeSainteAgathe:2019} and set tight Gaussian priors for two of the nuisance parameters that are less constrained by the data ($\beta_{HCD}$ and $b_{CIV}$). The choice of uninformative priors means that the shape of the posterior is given only by the likelihood. This means that any differences between Bayesian and frequentist results will be produced by the differences discussed in Section \ref{sec:theory}, and will not be influenced by prior choices.

\subsection{Fitting Methods}

Following the discussion in Section \ref{sec:theory}, we choose four fitting methods to compare over the population of synthetic data sets: 
\begin{enumerate}
    \item \textbf{Frequentist MLE}: The likelihood is maximised over all parameters and the uncertainties are given by the covariance around $\vec{\theta}_{best}$, using the Gaussian $\Delta\mathcal{L}_p$ values.
    \item \textbf{Frequentist scan}: A scan over \apar\ and \aperp\ using the profile likelihood, with the uncertainties given by confidence intervals which are set using MC simulations.
    \item \textbf{Bayesian scan}: A scan over \apar\ and \aperp\ using the profile likelihood, with the uncertainties given by credible regions, computed using Equation \ref{eq:bayes_scan}.
    \item \textbf{Bayesian sampler}: The full posterior distribution is computed, and the uncertainties on (\apar, \aperp) are given by credible regions after marginalization over the nuisance parameters.
\end{enumerate}

\lya\ BAO analyses have so far been frequentist, and used the first two methods. The Bayesian interpretation of the scan was also used whenever scan results were combined with other cosmological probes (e.g. \cite{Cuceu:2019}) as part of popular packages such as \texttt{CosmoMC} or \texttt{MontePython} \cite{Lewis:2002,Audren:2012,Brinckmann:2018}. However, the two interpretations of the scan were never tested together. This method is also a good middle ground between the frequentist scan and the Bayesian sampler, because the contrast with the first is only in the interpretation of uncertainty, while a comparison with the second allows us to directly test the profile likelihood.

The scan requires maximising the likelihood over all nuisance parameters at each point on the grid. As this operation can be performed independently for each point, we implemented a parallel version of the scan code in $\texttt{picca}$ to speed up our analysis. The frequentist analysis also requires a large number of MC simulations in order to compute the $\Delta\mathcal{L}^2_p$ as described in Section \ref{sec:theory}. We also implemented a parallel version of this step which is now available in $\texttt{picca}$.

In order to sample the full posterior distribution, we implemented an interface to the popular Nested Sampler \texttt{PolyChord}\footnote{Available at \url{https://github.com/PolyChord/PolyChordLite}} \cite{Handley:2015a,Handley:2015b} in \texttt{picca}. Nested samplers were designed to efficiently compute the Bayesian evidence \cite{Skilling:2004}, but they also provide accurate sampling of the posterior. In particular, nested samplers are very good at dealing with multimodal and highly degenerate posteriors. 


\subsection{Results}
\begin{figure}
    \centering
    \begin{subfigure}{.5\textwidth}
        \centering
        \includegraphics[width=1.0\textwidth,keepaspectratio]{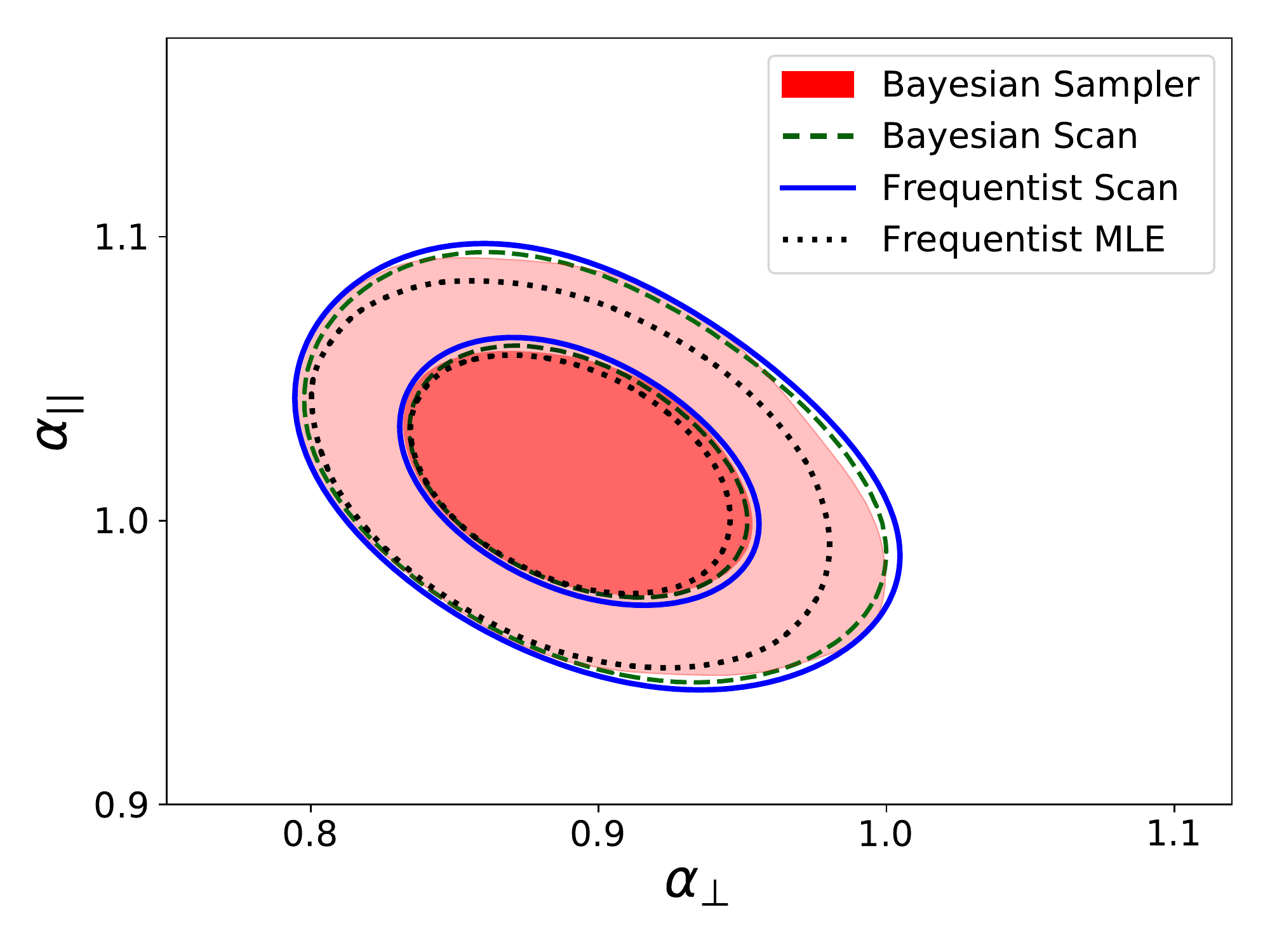}
    \end{subfigure}%
    \begin{subfigure}{.5\textwidth}
        \centering
        \includegraphics[width=1.0\textwidth,keepaspectratio]{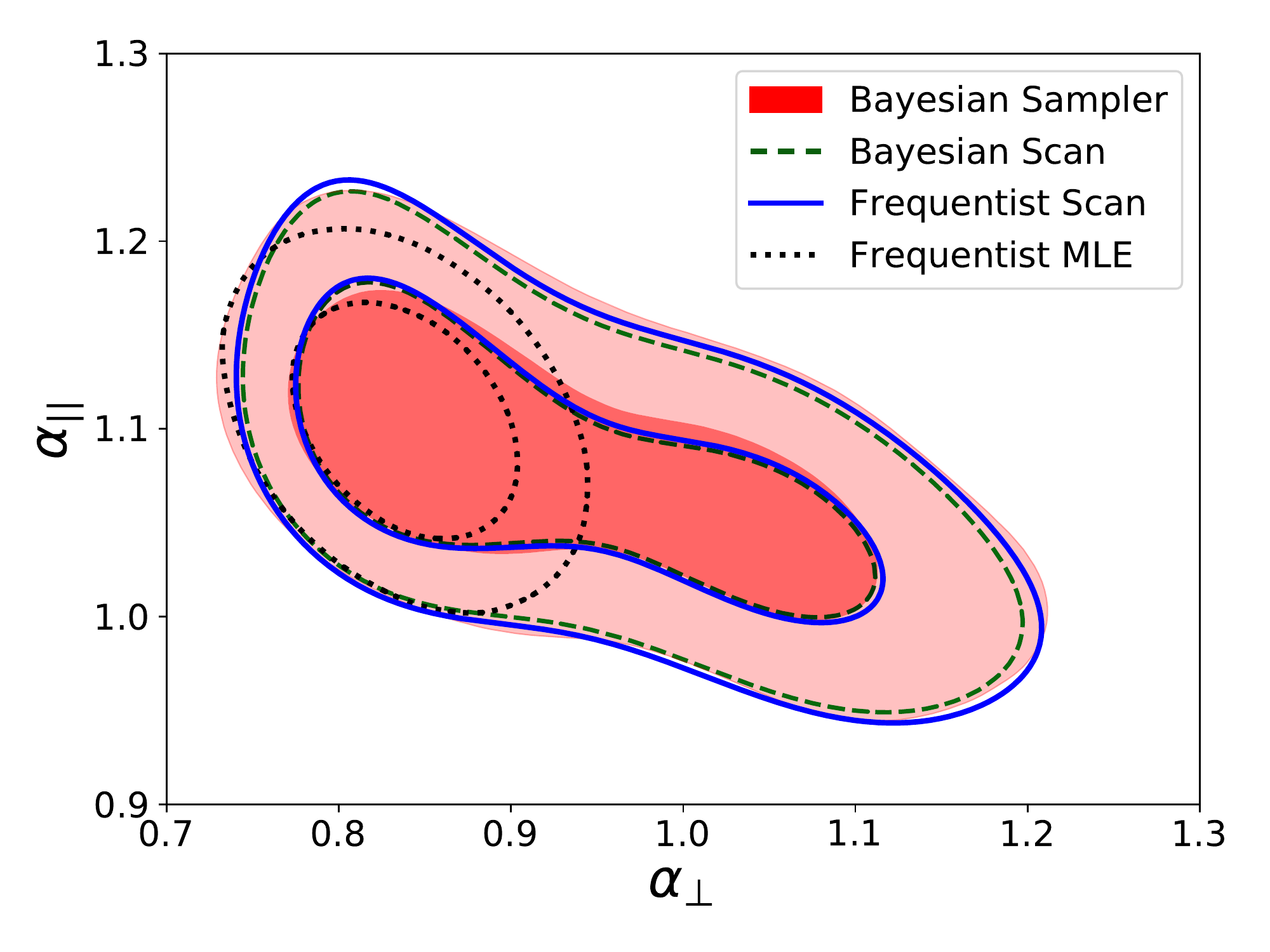}
    \end{subfigure}
    \caption{Comparison of BAO parameter constraints on mock correlation functions using the frequentist MLE and frequentist scan confidence intervals, and the Bayesian scan and Bayesian sampler credible regions. We showcase two mocks, one where the constraints are very close to Gaussian (left), and one where the constraints are strongly non-Gaussian (right). Note the different scaling. The last three methods produce remarkably similar results considering they use different methods and very different quantification of uncertainties. On the other hand, the frequentist MLE can fail to properly capture the uncertainty in the results, especially in non-Gaussian cases.}
    \label{fig:bao_contours}
\end{figure}

Noise on the correlation function can conspire to improve or smear the BAO peak. This means that a population of simulated correlation functions produced from the same covariance gives rise to a range of possible BAO constraints, from very tight, closely Gaussian fits, to very degenerate, multimodal ones where the peak is barely detected. In Figure \ref{fig:bao_contours}, we showcase two examples of results from this population. The one on the left is one of the best BAO constraints in the population, and is very close to Gaussian. On the other hand, the one on the right gives one of the worst constraints and is very non-Gaussian. We note that in the case of DR14-like errors on the correlation function, roughly $70\%$ of constraints (computed using visual inspection of the mock fits) are close to Gaussian, similar to the one on the left. On the other hand, $4\%$ are catastrophic failures, where the constraints are multimodal and strongly non-Gaussian in both parameters (BAO not detected).

Results of the four methods described above are presented in Figure \ref{fig:bao_contours}. The last three methods agree very well with each other even in the non-Gaussian case. This agreement is remarkable, because as discussed in Section \ref{sec:theory}, they answer fundamentally different questions. The frequentist scan uses confidence intervals to quantify uncertainty, while the two Bayesian methods use credible regions. The agreement between the frequentist scan and the Bayesian scan shows us that BAO results are robust to different interpretations of uncertainty. On the other hand, the agreement between the Bayesian scan and the Bayesian sampler shows us that the profile likelihood is a good approximation for fitting the BAO peak from the \lyaf\ correlation function. 

The Frequentist MLE, which assumes Gaussianity, does not fare as well. It produces results that are close to the ones obtained using the other three methods in the Gaussian mock, although slightly smaller. On the other hand, it completely fails to capture the uncertainty in the results for the non-Gaussian mock. As these non-Gaussian posteriors make up roughly $30\%$ of the population of possible results, this approximation is inadequate for dealing with current data sets.

The comparison of specific posterior results such as in Figure \ref{fig:bao_contours} is Bayesian in nature. However, we can also compare them using a frequentist approach by computing the fraction of the interval population that contains the truth $\vec{\theta}_{true}$ within the $68\%$ and $95\%$ regions. We find that the frequentist scan, the Bayesian scan and the Bayesian sampler are again in remarkable agreement, with roughly $55-60$ mocks containing the truth within their $68\%$ regions, and roughly $92$ of them containing the truth within their $95\%$ regions. On the other hand, the MLE fails this test as well, with only $44$ of the $68\%$ intervals, and $74$ of the $95\%$ intervals containing the truth. These numbers are affected by sample noise because we only have $100$ simulations, and as such are very rough. In particular, if the profile likelihood works perfectly, by construction the fractions for the frequentist scan will tend to $68\%$ and $95\%$ for a large number of samples.

These results do, of course, rely on the large uniform priors assumed for most parameters. Non-uniform or overly restrictive priors would have an impact on our conclusions as some of the differences could come from the choice of priors. In a study similar to our work, \cite{Chan:2018} arrive at a different conclusion when tight priors are used on the BAO parameters.  We discuss the similarities and differences between our study and \cite{Chan:2018} in Appendix \ref{sec:app_comp}.

Another relevant question that distinguishes these methods is that of the computational cost. MLE is by far the fastest method, with computation times of order $10^{-1}-10^0$ CPU hours. However, as we just showed, MLE alone is inadequate for current \lyaf\ BAO analyses. For the scan we find computation times of order $10^2 - 10^4$ CPU hours using typical grid sizes of $30\times30$ up to $50\times50$. Furthermore, the frequentist interpretation of the scan requires a large set of MC simulations to compute the $\Delta\mathcal{L}_p$ necessary for setting the right CIs. In the best case these require an extra $10^{3}$ CPU hours for $10000$ MC mocks. \texttt{PolyChord} with a typical setup ($nlive = 25\times $ number of parameters, $\mathit{num repeats} = 3\times $ number of parameters) performs over a timescale similar to the scan ($10^2 - 10^4$ CPU hours), and not only computes the full posterior distribution, but also its integral (the Bayesian evidence).




\section{The Full Posterior of eBOSS DR14}
\label{sec:data}

We now turn our attention to the full posterior distribution. As we have shown above, when the only interest is measuring the BAO peak position, both the scan and the sampler produce very similar results. However, the sampler also computes accurate distributions for all the other parameters. This wealth of information is generally ignored. Some of these are astrophysical parameters, such as the bias of high column density (HCD) absorbers ($b_{HCD}$). These parameters tend to be very correlated with each other and have proved to be very sensitive to modelling choices, which makes their measurement less robust compared to that of the BAO peak.

The computation of the full posterior distribution allows us to access this previously ignored information. It allows us to study the complex high-dimensional distribution of these parameters, and to better understand how our modelling choices affect their measurement. Furthermore, our use of the Nested Sampler \texttt{PolyChord} means that we can deal with very degenerate and even multimodal distributions. This sampler also computes the Bayesian evidence, thus providing an accurate tool for model comparison even for strongly non-Gaussian posteriors. As such, it allows us to test possible extensions to the baseline model that may have previously appeared daunting due to their complex interaction with other parameters.

\subsection{The Baseline Analysis}
\label{subsec:standard}
\begin{center}
    \begin{table}
            \centering
            \begin{tabular}{c p{70mm} c}
                \hline\hline            
                Parameter & Description & Prior \cr 
                \hline
                \apar, \aperp & BAO peak position & $\Pi[0.1,2]$ \cr
                $b_{\eta Ly\alpha}$ & \lya\ velocity bias & $\Pi[-0.5,0]$ \cr
                $\beta_{Ly\alpha}$ & \lya\ RSD parameter & $\Pi[0.1,5.0]$ \cr
                $\beta_{QSO}$ & Bias of HCDs & $\Pi[-0.2,0]$ \cr
                $\Delta r_{||} [h^{-1}Mpc]$ & Shift due to QSO redshift errors & $\Pi[-10,10]$ \cr
                $\sigma_v [h^{-1}Mpc]$ & Smoothing parameter for QSO non-linear velocities and redshift precision & $\Pi[2,15]$ \cr
                $\xi^{TP}_0$ & Amplitude parameter of quasar radiation & $\Pi[0,2]$ \cr
                $b_{HCD}$ & Bias of HCDs & $\Pi[-0.2,0]$ \cr
                $\beta_{HCD}$ & RSD parameter of HCDs & $\mathcal{N}(0.5, 0.2^2)$ \cr
                $b_{\eta CIV(eff)}$ & Velocity bias of metal absorber & $\mathcal{N}(-0.005, 0.0026^2)$ \cr
                $b_{\eta SiII(1190)}$ & Velocity bias of metal absorber & $\Pi[-0.2,0]$ \cr
                $b_{\eta SiII(1193)}$ & Velocity bias of metal absorber & $\Pi[-0.2,0]$ \cr
                $b_{\eta SiIII(1207)}$ & Velocity bias of metal absorber & $\Pi[-0.2,0]$ \cr
                $b_{\eta SiII(1260)}$ & Velocity bias of metal absorber & $\Pi[-0.2,0]$ \cr
                \hline
            \end{tabular}
            \caption{Sampled parameters and their prior limits. We use flat priors $\Pi[a,b]$ for most parameters, with limits $a$ and $b$ chosen such that the prior is uninformative. Following \cite{DeSainteAgathe:2019,Blomqvist:2019}, we use Gaussian priors $\mathcal{N}(\mu, \sigma^2)$ with mean $\mu$ and standard deviation $\sigma$ for $\beta_{HCD}$ and $b_{\eta CIV(eff)}$,.}
        
            \label{tab:pars}
    \end{table}
\end{center}

\begin{figure}
    
    \includegraphics[width=1.0\textwidth,keepaspectratio]{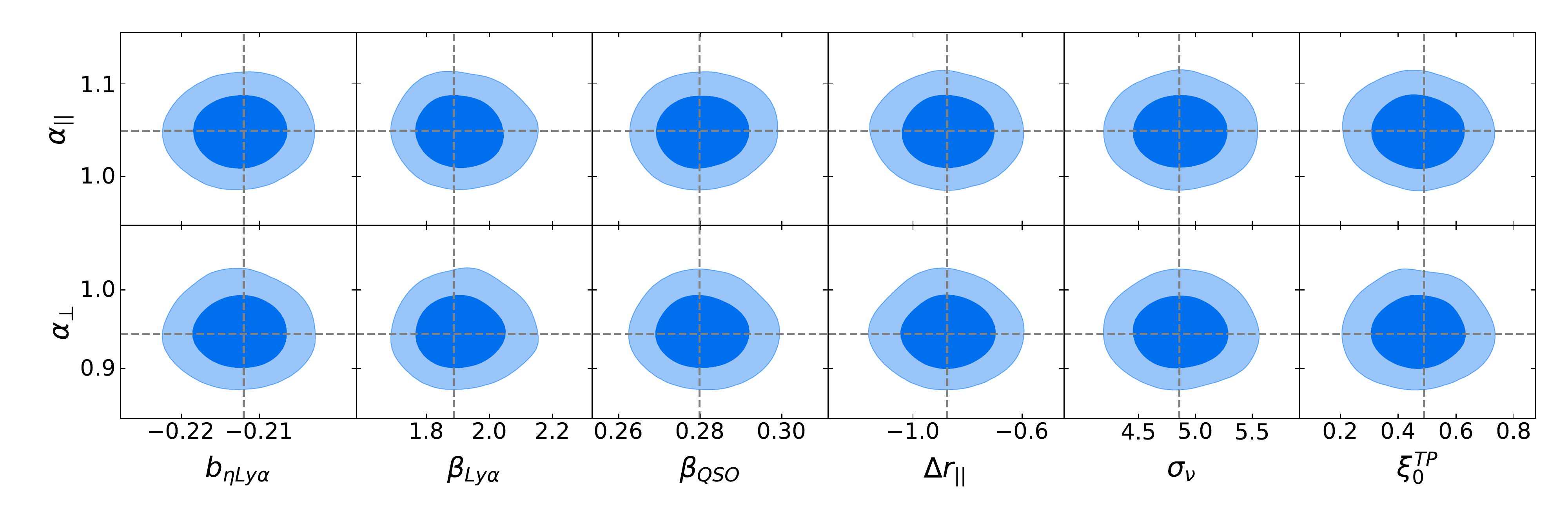}%
    
    \includegraphics[width=1.0\textwidth,keepaspectratio]{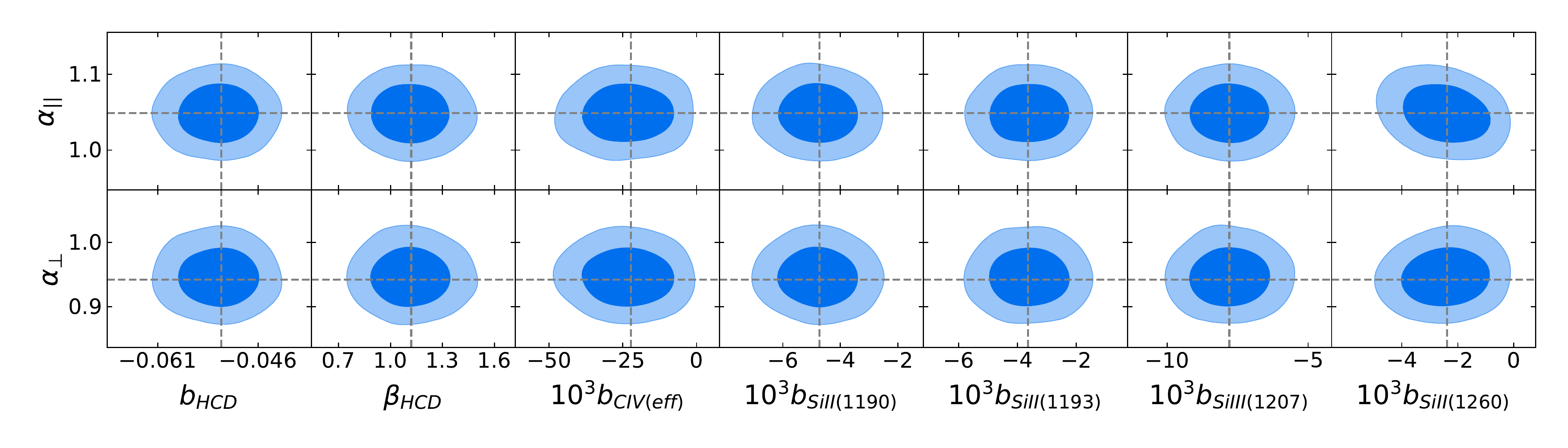}%
    
    \caption{Projected posterior distributions of the BAO parameters (\apar, \aperp), versus the other model parameters, using the full eBOSS DR14 \lyaf\ data. There are no major correlations between \apar$/$\aperp\ and any of the other parameters, which shows the robustness of BAO parameters to different modelling choices.}
    \label{fig:bao_corr}
    
\end{figure}    

We start by using \texttt{PolyChord} to analyse all \lyaf\ eBOSS DR14 correlation functions. In contrast to the last section, we include the correlation of \lya\ absorbers in the \lya\ region with \lya\ absorbers in the \lyb\ region, $Ly\alpha(Ly\alpha)\times Ly\alpha(Ly\beta)$, and the cross-correlation with quasars \lyaqso. We use the models presented in \cite{DeSainteAgathe:2019} and \cite{Blomqvist:2019} to model the auto-correlation and the cross-correlation with quasars respectively. Our analysis follows the same steps, but we do not include relativistic effects, and we use only one parameter to model the bias of HCDs (instead of three). As in the previous section, we use broad flat priors on all parameters (except $\beta_{HCD}$ and the bias of foreground CIV absorption ($b_{CIV}$) which have Gaussian priors). The model parameters used and their priors are presented in Table \ref{tab:pars}.

In Figure \ref{fig:bao_corr} we show the projected posterior distributions of \apar\ and \aperp\ versus the other parameters. This shows the robustness of BAO measurements as none of the parameters are correlated with (\apar, \aperp). The only exception is the bias of the $SiII(1260)$ absorbers ($b_{SiII(1260)}$) which has a small correlation with \apar. This is due to the line causing an increased correlation along the line of sight at a separation of $r_{||} \approx 105 \textrm{h}^{-1}\textrm{Mpc}$, which is very close to the BAO peak. However, this metal contamination is barely detected (at $\sim2\sigma$).



\subsection{A simple extension}
\label{subsec:extension}

We now use \texttt{PolyChord} to illustrate the advantages of sampling the full posterior. We do this by sampling $L_{HCD}$ as an extension to the baseline analysis. This is a parameter used to describe the contamination by HCDs and it corresponds to a typical length scale. HCDs with a length scale above $\sim 14\textrm{h}^{-1}\textrm{Mpc}$ are efficiently identified and masked before computing the correlation function. This means unidentified systems are expected to have a typical length scale below this value, but above the bin width of the correlation function ($4\textrm{h}^{-1}\textrm{Mpc}$). This parameter has so far been fixed to a value of $L_{HCD} = 10 \textrm{h}^{-1}\textrm{Mpc}$ following the study by \cite{Rogers:2018}. Different values for this parameter were tested by \cite{DeSainteAgathe:2019,Blomqvist:2019} to confirm there are no biases when measuring the BAO position, however, completely freeing this parameter proved challenging for the minimiser to deal with. We set a flat prior given by: $\Pi[2,30] \textrm{h}^{-1}\textrm{Mpc}$. 

\begin{figure}
    \centering
    \includegraphics[width=1.0\textwidth,keepaspectratio]{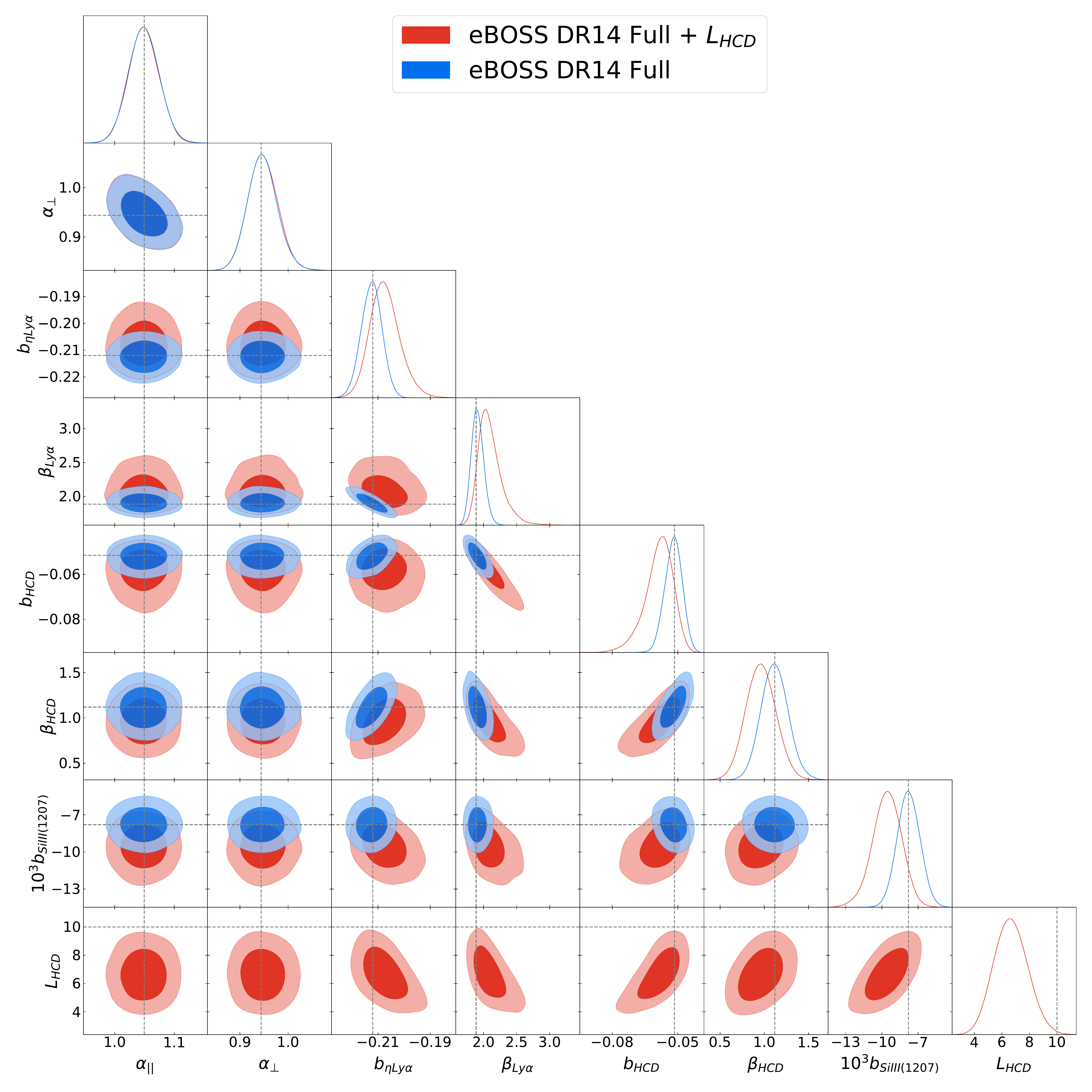}
    \caption{Triangle plot showing projected posteriors of DR14 \lyaf\ results using the baseline analysis versus an extension where the typical scale of high column density absorbers, $L_{HCD}$, is sampled. The best fit results for the baseline analysis are given by the dashed lines. The first two columns show that the measurement of the BAO scale is very robust to this change. However, the other parameters plotted are correlated with $L_{HCD}$, and as such, their posteriors are significantly affected by this parameter.}
    \label{fig:lhcd}
\end{figure}

As the full projected posterior is too large to show, we chose a subset of parameters and plot their posteriors in Figure \ref{fig:lhcd}. We show \apar\ and \aperp\ to study the impact of $L_{HCD}$ on measurements of the BAO peak position. From the other parameters we chose those whose posterior is affected by $L_{HCD}$. We also plot the posterior using the basic model where $L_{HCD}$ is fixed for comparison. Plots of the full posterior are available as part of a Jupyter Notebook at \url{https://github.com/andreicuceu/eBOSS-Lya-Posteriors}.

The first two columns in Figure \ref{fig:lhcd} show that even though the posterior is significantly affected by the choice of $L_{HCD}$, the BAO peak position is very robust to it. In particular, the $\alpha_{||} - \alpha_{\bot}$ posterior of the extended model is in excellent agreement with the results from the baseline analysis. $L_{HCD}$ is constrained within the expected region: $L_{HCD} = 6.6 \pm 1.2,\;^{+2.5}_{-2.3} \; \textrm{h}^{-1}\textrm{Mpc}\; (68\%,95\%)$. As can be seen in the bottom row, the other parameters shown in Figure \ref{fig:lhcd} are correlated with $L_{HCD}$, and as such their posterior is significantly affected. The constraints on the biases and RSD parameters of the \lyaf\ and HCDs are all wider than in the baseline analysis, showing the fragility of these parameters to modelling choices. Interestingly, the bias of the SiIII(1207) absorption is also very correlated with $L_{HCD}$.

Comparisons such as the one presented in Figure \ref{fig:lhcd} are beneficial because they visually show us how model parameters behave, and how our modelling choices influence our results.



\section{Conclusions}
\label{sec:conclusion}

Analyses of the \lyaf\ 3D correlation function have so far focused on measuring the position of the BAO peak at the expense of analyses of other parameters and their interaction. A frequentist methodology has been used so far, despite the BAO results being subsequently used as part of Bayesian analyses when combined with other probes. In this work, we performed for the first time a Bayesian analysis of the \lyaf\ correlation function, and we computed the full posterior using the Nested Sampler \texttt{PolyChord}. 


We started by discussing the different approaches to fitting the correlation function in Section \ref{sec:theory}, and in particular we focused on the methods used and the difference in the quantification of uncertainty. Frequentists use Confidence Intervals computed from the population of possible data sets, while Bayesians use credible regions computed from the posterior distribution of the parameters given the data and the prior. Furthermore, when dealing with high-dimensional model spaces, Bayesians have access to tools such as MCMC. On the frequentist side, approximations such as the profile likelihood are used. A scan is performed over a few parameters (usually anything above 2 or 3 is computationally infeasible), and the likelihood is maximised over other parameters at each point. This is generally a good approximation, however, it can fail in some cases, such as in the toy example presented in Section \ref{sec:theory}. 

We compared the different methodologies on synthetic correlation functions of eBOSS DR14 in Section \ref{sec:mocks}. We used both a Bayesian and a frequentist interpretation of scan results. This allows us to test potential differences from the quantification of uncertainty by comparing the two. We can also test the profile likelihood approximation by comparing the Bayesian scan with Bayesian sampler results. We showed that the three methods agree remarkably well on both a mock with a tight BAO constraint, and one with a large non-Gaussian one. We also plotted results using the frequentist Maximum Likelihood Estimation, where the uncertainties are set using the second derivative around the peak likelihood. This method fails to capture the uncertainty in non-Gaussian cases. Considering these make up roughly $\sim 30\%$ of the population, the MLE is inadequate for current data sets.

By computing the Bayesian interpretation of the scan and showing it agrees very well both with the frequentist scan and the Bayesian sampler, we have confirmed that scan results can safely be combined with other probes as part of Bayesian packages commonly used in cosmology.

In Section \ref{sec:data} we turned our attention to the full posterior distribution of eBOSS DR14 \lyaf\ correlation functions. We showed that the BAO peak position parameters, \apar\ and \aperp, do not have any strong correlation with any of the other parameters, and as such, they are very robust to modelling choices. We extended the baseline analysis by sampling the typical length scale of HCD systems, which was previously fixed to $L_{HCD} = 10\textrm{h}^{-1}\textrm{Mpc}$. We compared the projected posteriors of the two models, and showed that it has no impact on the measured BAO position. We measured $L_{HCD} = 6.6 \pm 1.2 \; \textrm{h}^{-1}\textrm{Mpc}$ ($68\%$), and find that this parameter is correlated with many other astrophysical parameters and as such has a significant impact on their posteriors. Plots of the full posterior are available at \url{https://github.com/andreicuceu/eBOSS-Lya-Posteriors}.

Comparisons such as the one presented in Section \ref{subsec:extension} are now easy and fast to perform. Furthermore, this benefit does not come at a major computational cost. As discussed in Section \ref{sec:mocks}, \texttt{PolyChord} requires similar computational time compared to a 2D scan using the profile likelihood. Beyond constraining the BAO peak parameters, \texttt{PolyChord} also computes the full posterior distribution of all parameters and the Bayesian evidence.

Our improvements are freely available as part of the community code \texttt{picca}. We hope that these tools will be used to improve future \lyaf\ analyses by using the full posterior distribution to study complex parameter distributions and inform modelling choices. 


\acknowledgments

We thank Lorne Whiteway, Pablo Lemos, An\v{z}e Slosar, Ashley Ross and Kwan Chuen Chan for their comments and advice on the draft. This work was partially enabled by funding from the UCL Cosmoparticle Initiative. AC was supported by a Science and Technology Facilities Council (STFC) studentship. AFR acknowledges support by an STFC Ernest Rutherford Fellowship, grant reference ST/N003853/1 and further support by STFC Consolidated Grant number ST/R000476/1.

\bibliographystyle{JHEP.bst}
\bibliography{main}

\appendix

\section{Computation of CIs}
\label{sec:app_chisq}

If the model is not a linear function of the parameters, then the $\Delta\mathcal{L}_p$ values, necessary for setting confidence intervals from grid results, usually cannot be computed analytically. In general these values must be computed from the population of the possible realisations of the data. However, this population is inaccessible if we can only perform the experiment once, and we must rely on simulated populations of data sets instead of the real one \cite{Press:2007}. In cosmology this could be achieved by running the full analysis many times on simulations, however as these are extremely expensive computationally, we usually rely on Monte Carlo (MC) simulations instead. The idea behind MC is to use the measured correlation functions and their covariance matrices together with a model of the data to produce random realisations of our data. See Chapter 15.6 of \cite{Press:2007} for a detailed discussion of this algorithm. Note that the \chisq\ is generally used instead of the log-likelihood, and $\Delta\chi^2$ values are computed, however, we use the log-likelihood to maintain consistency and because the two are proportional if the likelihood is Gaussian: $\chi^2 \propto -2 \log \mathcal{L}$. 

We want to produce samples from the distribution of possible data sets given the true model and the true parameter values: $P(D|\vec{\theta}_{true},\mathcal{M}_{true})$. However, finding this distribution is the goal of our experiment, and as such it is inaccessible at this point. Therefore, we start by assuming that the shape of $P(D|\vec{\theta}_{true},\mathcal{M}_{true})$ is similar to the shape of $P(D|\vec{\theta}_0,\mathcal{M}_0)$, where we are using our particular model $\mathcal{M}_0$, and the best fit parameters given that model and our data: $\vec{\theta}_0$. This means that we can use $\vec{\theta}_0$ as a surrogate for $\vec{\theta}_{true}$ and create a population of synthetic data sets from a multivariate normal with mean $\mathcal{M}_0(\vec{\theta}_0)$ and covariance matrix given by the measurement errors of the data (again assuming our data is normally distributed).

For each mock data set created using this method we find the best fit parameter values and call them $\vec{\theta}_\alpha$. If we compute the quantity $\Delta\mathcal{L} \equiv \mathcal{L}(\vec{\theta}_0) - \mathcal{L}(\vec{\theta}_\alpha)$ for each mock data set, we obtain a chi-square distribution with $n$ degrees of freedom, where $n$ is the number of parameters. By analysing a large population of mocks and computing this quantity, the $\Delta\mathcal{L}_p$ is easily obtained by looking at the $\Delta\mathcal{L}$ value that contains the fraction of mocks corresponding to $p$ for that chi-square distribution. 

If we are only interested in a subset of $\nu$ parameters, then the quantity of interest is $\Delta\mathcal{L}_{\nu} \equiv \mathcal{L}_{\nu} - \mathcal{L}(\vec{\theta}_\alpha)$, where $\mathcal{L}_{\nu}$ is found by fixing the parameters of interest to their best fit values from $\vec{\theta}_0$ and maximising over the other parameters for each mock. $\Delta\mathcal{L}_{\nu}$ is a chi-square distribution with $\nu$ degrees of freedom, and the same procedure as above is used to find $\Delta\mathcal{L}_p$. 

\section{Comparison with Chan et al. 2018}
\label{sec:app_comp}

The Dark Energy Survey (DES) collaboration performed a study similar to the one presented in this work, testing different fitting methods with the aim of measuring BAO using angular correlation functions in tomographic bins \cite{Chan:2018}. They also compared the performance using an MLE approach, the profile likelihood and an MCMC. When they computed the mean and uncertainty from the profile likelihood they used the mean and variance across the scan weighted by the likelihood value at each point. This is in contrast with our method of finding constant $\chi^2$ surfaces, and as such it may lead to different results.

The main difference is that they worked with isotropic BAO which means they only need to measure one parameter, $\alpha$. Furthermore, they also pruned their population of synthetic data sets to those where the $68\%$ $\alpha$ constraints are entirely contained within the prior range $[0.8,1.2]$ (where $\alpha = 1$ is the true value). The result is that the remaining mocks are the ones where the BAO peak has a $\Delta \chi^2 =1$ region within their prior range on $\alpha$, which ensures the posterior distribution of $\alpha$ can be approximated as Gaussian. They found that all three methods perform very well, and after investigating the population statistics they conclude that the MLE is the best tool in these conditions.

In the particular case of isotropic BAO we find similar results. As the model has less freedom, most constraints are Gaussian and all three methods work very well. However, they found that the MCMC has a larger bias in estimating $\alpha$ compared to the MLE (although both are very small). We believe that this is because they used the median to quantify the MCMC result, while we are using the maximum posterior point. The best fit point given by an MCMC (assuming flat priors) and MLE should be the same if both are run appropriately, and as such there should not be any difference in the bias of the peak for a unimodal posterior.

In Appendix A, \cite{Chan:2018} worked with a larger range ($[0.6,1.4]$) and showed that in this case the MCMC is the best approach, with both the MLE and profile likelihood showing small biases. This is in line with our results, considering that in this case there are some non-Gaussian results that are better fit using an MCMC approach. However, we also found the profile likelihood still works very well in these cases. This difference in results could be either caused by the difference in quantifying uncertainty as discussed above, or by a failure of the profile likelihood approximation when dealing with the nuisance parameters.

\end{document}